\documentstyle[amsfonts,twocolumn,prl,aps,epsf]{revtex}

\begin{document}
\draft
\title{Viscoelasticity near the gel-point: a molecular dynamics study}

\author{D.C. Vernon, Michael Plischke}
\address{Physics Department, Simon Fraser University,
         Burnaby, British Columbia, Canada V5A 1S6}
\author{B\'{e}la Jo\'{o}s}
\address{Ottawa Carleton Institute of Physics, University of Ottawa Campus,
Ottawa, Ontario, Canada K1N-6N5}
\date{\today}
\maketitle
\begin{abstract}
We report on extensive molecular dynamics simulations on systems of
soft spheres of functionality $f$, {\em i.e.} particles that are
capable of bonding irreversibly with a maximum of $f$ other
particles. These bonds are randomly distributed throughout the system
and imposed with probability $p$. At a critical concentration of
bonds, $p_c\approx 0.2488$ for $f=6$, a gel is formed and the shear
viscosity $\eta$ diverges according to $\eta\sim (p_c-p)^{-s}$. We
find $s \approx 0.7$ in agreement with some experiments and with a
recent theoretical prediction based on Rouse dynamics of phantom
chains. The diffusion constant decreases as the gel point is
approached but does not display a well-defined power law.
\end{abstract}
\narrowtext The behavior of transport coefficients and elastic moduli
near the gelation transition has been discussed in the literature for
many years\cite{adam96}. To date no consensus on either the
theoretical or experimental side has emerged as far as the critical
behavior of these quantities is concerned. The phenomenology is as
follows. As monomers or polymers are randomly crosslinked to each
other in a melt, the shear viscosity $\eta$ increases with crosslink
concentration $p$ and diverges at a critical concentration $p_c$ at
which an amorphous rigid network is formed.  Experiment and theory
both yield $\eta\sim(p_c-p)^{-s}$ but there is no general agreement
regarding the value of the exponent $s$. Indeed there is good reason
to expect at least two different universality classes: As de Gennes
showed\cite{degen77}, vulcanization (crosslinking of very long chains)
must be distinguished from gelation (crosslinking of short chains or
monomers) as far as critical behavior is concerned. This conclusion is
supported by recent experiments\cite{colby99,colby95} which show quite
clearly that chain length is a relevant parameter.

Before describing our model and calculations, we discuss briefly the
experimental and theoretical situation for gels putatively in the
percolation (short chains) universality class. One group of
experiments has produced exponent values for the shear viscosity in
the range $0.6\leq s\leq 0.9$\cite{adam,durand}. Another group
\cite{martin1,martin2,colby95} has reported values for $s$ in the
range 1.1 --- 1.3 and interpreted \cite{colby95} these in terms of a
Rouse model without hydrodynamics. On the theory side, de
Gennes\cite{degen79} argued that the viscosity is analogous to the
conductivity of a random mixture of normal conductors and
superconductors, with an exponent $s\approx 0.67$. The aforementioned
Rouse model\cite{martin2} prediction is $s=2\nu-\beta\approx 1.35$,
where $\nu\approx 0.88$ and $\beta\approx 0.41$ are the correlation
length and order parameter exponents of percolation theory in three
dimensions. Finally, a recent theoretical analysis of a different
model with Rouse-dynamics\cite{zipp99} has predicted
$s=\phi-\beta\approx 0.7$ where $\phi\approx 1.11$ is the crossover
exponent of a random resistor network. Given this wide disparity in
both theoretical and experimental results, computer simulations may
help to clarify the situation.

The shear modulus $\mu$ of a rigid network near the gel point is
typically entropic in nature and it vanishes with a power law of its
own as the gel point is approached from the rigid phase $\mu\sim
(p-p_c)^t$. Recent numerical work on systems in the percolation
universality class\cite{MP99} has provided evidence that $t=f$ in both
two and three dimensions, where $f$ is the exponent that describes the
critical behavior of a randomly disordered network of conductors and
insulators near the percolation point. This result is consistent with
another argument of de Gennes\cite{degen76}. In the dynamical scaling
theory of the gelation transition, the two exponents $t$ and $s$ are
not independent, but rather obey the sum rule $s+t=z$ where $z$
describes the divergence of the longest relaxation time in the
incipient gel: $t^\ast=t_0(p_c-p)^{-z}$\cite{adam96}. This connection
allows an important consistency check between the results reported
here and those of \cite{MP99}.

The model that we simulate is
capable of describing the entire range from simple liquid to entropic
solid. All particles interact through a soft sphere potential
$V_{sa}(r_{ij})=\epsilon(\sigma/r_{ij})^{36}$ \cite{ph00} with
$\sigma=1$ and, for our simulations, $k_BT/\epsilon=1$. If there are
no other interactions, this system forms a simple three-dimensional
liquid at least at low density. All of our simulations are done at a
volume fraction $\Phi=\pi N\sigma^3/6V=0.4$ which is well away from
the liquid-solid coexistence density.  The viscosity of the system is
progressively increased by introducing random crosslinks between
particles. Specifically, the system of particles is initially placed
on a simple cubic lattice that fills the cubic computational box. Each
particle may bond with probability $p$ with each of its six nearest
neighbors.  This bonding is permanent and enforced by the spherically
symmetric potential $V_{nn}(r_{ij})=\frac{1}{2}k(r_{ij}-r_0)^2$ with
$k=5\epsilon/\sigma^2$ and $r_0=(\pi/6\Phi)^{1/3}\sigma$.  For
$p<p_c$, where $p_c\approx 0.2488$ is the bond percolation probability
of the simple cubic lattice, the system consists of finite clusters of
varying masses. For $p>p_c$ the system is an entropic solid with
nonvanishing shear modulus. This system has been previously studied by
us\cite{MP99} for $V_{sa}=0$ and $p>p_c$. Farago and Kantor and Cohen
and Plischke\cite{FK00} have shown that self-avoidance is irrelevant
as far as the critical behavior of the elastic constants is
concerned.

In the simulations, the system of particles is first
equilibrated for $5\times 10^4$ time steps with Brownian dynamics. At
the end of this equilibration time, the damping and thermal noise are
turned off and the subsequent evolution is conservative. The equations
of motion are integrated with a standard velocity Verlet algorithm
\cite{allen} with a time step of $\delta
t=0.005\sqrt{m\sigma^2/\epsilon}$. The shear viscosity $\eta(p)$ and
the self-diffusion constant $D(p)$ are then calculated from the
appropriate Green-Kubo formula \cite{hansen}:
\begin{eqnarray}\label{eta}
\eta&=&\lim_{t_{max}\to\infty}\int_0^{t_{max}}dt C_{\sigma\sigma}(t)
\nonumber\\ &=&\lim_{t_{max}\to\infty}\frac{1}{3Vk_BT}\int_0^{t_{max}}
dt\sum_{\alpha<\beta}\left\langle\sigma^{\alpha\beta}(t)\sigma^{\alpha\beta}(0)
\right\rangle
\end{eqnarray}
where
\[
\sigma^{\alpha\beta}=\sum_{i=1}^Nmv_{i\alpha}v_{i\beta}-\sum_{i<j}\frac{r_{ij\alpha}
r_{ij\beta}}{r_{ij}}V'(r_{ij})
\]
and where $V'$ is the derivative of the total potential
energy. Similarly, the diffusion constant is given by the familiar
expression:
\begin{equation}\label{D}
D=\lim_{t_{max}\to\infty}\frac{1}{3N}\sum_{i=1}^N\int_0^{t_{max}}
dt\left\langle{\bf v}_i(t)\cdot{\bf v}_i(0)\right\rangle\,.
\end{equation}
It is well known that the velocity-velocity correlation function
decays extremely slowly, typically with a `long time tail' $t^{-3/2}$
power law, even in simple liquids \cite{hansen}. We find the same
long-time behavior in our simulations as well, remarkably for all
values of $p$.  There is considerably more disagreement regarding the
stress-stress correlation function. Extended mode-coupling theory
\cite{bal90} suggests that close to the melting point the stress
correlator decays exponentially at long times. Powles and Heyes
\cite{ph00} have found that both an exponential decay and a Lorentzian
provide a reasonably good fit to their data in the dense liquid
regime.  In our situation where clusters of various sizes form the
system, the decay is dramatically affected by crosslinking and becomes
very slow close to the gel-point. Therefore in the evaluation of
(\ref{eta},\ref{D}) we have used time series from $t_{max}=300\tau$ to
$t_{max}=1200\tau$ where $\tau$ is the average time between collisions
for particles in the uncrosslinked liquid ($p=0$). Even with such long
runs, for $p$ close to $p_c$ an estimate of the residual integral to
$t=\infty$ had to be added. This is discussed further below.

\begin{figure}[t]
\centerline{\epsfxsize=6.5cm 
\epsfbox{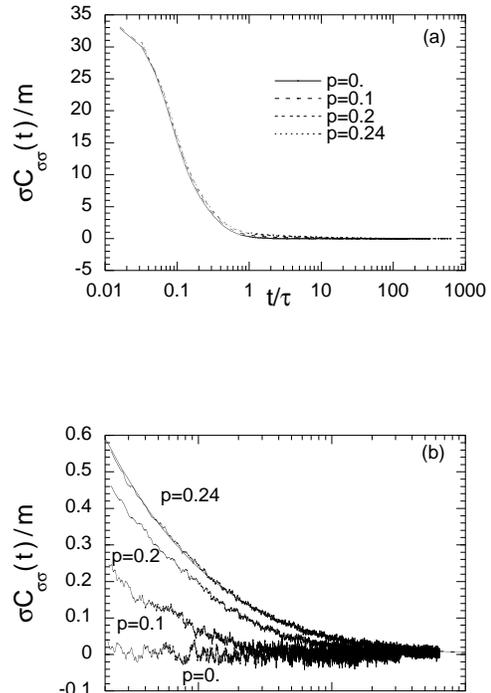}}
\caption{The dimensionless stress-stress correlation function $\sigma
C_{\sigma\sigma}/m$ for
$L=12$ and several crosslinking probabilities for (a) all $t/\tau >0$ and (b) for
$t/\tau >2.$ In the uppermost curve in (b) we also show the fit of
$C_{\sigma\sigma}$ to a stretched exponential (solid line). For $p=0.1$ the data
are obtained from 30 different crosslinkings, for larger $p$ from between 100
and 200 crosslinkings.}
\label{figsig}
\end{figure}

We have simulated systems consisting of $N=L^3$ particles with $L=5, 8,
12$ and $20$, the first three over the concentration range $0\leq
p\leq0.24$, the fourth only for $p\geq 0.20$. For these relatively
small systems, the probability that one of the clusters percolates in
at least one of the three directions is an issue. If there is
percolation, then the viscosity is not a well-defined quantity and the
sample is characterized instead by a shear modulus. Therefore, we have
eliminated all percolating samples from the calculation. For $L=5$, a
non negligible number of samples percolates in at least one direction
at $p=0.15$; at $L=20$ percolation becomes significant only at
$p=0.23$. Figure \ref{figsig} depicts the stress correlator for
$N=12^3$ particles for a set of crosslink concentrations. The top
panel shows this function over the entire range in time, the lower
panel for $t>2\tau$. The top panel shows that the changes for short
times between the liquid ($p=0$) and the incipient gel ($p=0.24$) are
extremely small. The effect of increasing crosslink density is
illustrated in the lower panel where it is clearly seen that the decay
of the correlation function becomes progressively slower as the gel
point is approached and that even at the longest time, for $p\geq 0.2$
the correlator is nonnegligible. Therefore, integrating
$C_{\sigma\sigma}(t)$ only to $t_{max}$ would result in an
underestimate of the shear viscosity. In order to capture the
remaining contribution, we have fit $C_{\sigma\sigma}(t)$ with a
stretched exponential over various windows $(t_s,t_{max})$ for
starting values $t_s>2\tau$. Such a fit is shown in Figure
\ref{figsig}b for the uppermost curve ($p=0.24$) over the range $2
\tau\leq t\leq t_{max}$. The fit is essentially indistinguishable from
the simulation. However, there is no fundamental reason to believe
that a stretched exponential is the functional form of the long-time
behavior of $C_{\sigma\sigma}(t)$\cite{ladd87}. This is the reason
that we have chosen several different starting points for the fit: the
spread in values of the remaining integral of the fitting function from
$t_{max}$ to infinity provides an estimate of the error associated
with this part of the calculation.

\begin{figure}
\centerline{\epsfxsize=6.5cm 
\epsfbox{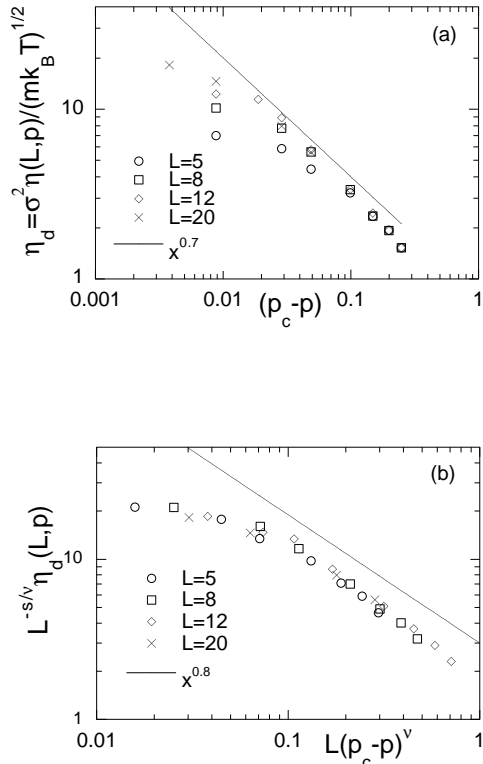}}
\caption{Log-log plot of the dimensionless shear viscosity as function of
crosslinking probability $p$: (a) raw data; (b) finite-size scaling form of the
data.}
\label{figeta}
\end{figure}

In Figure \ref{figeta} we
display our data for the dimensionless shear modulus
$\eta\sigma^2/\sqrt{mk_BT}$ in two versions. In part (a) the raw data
is shown as function of $(p_c-p)$ together with a guide to the eye of
the form $a(p_c-p)^{0.7}$. This function clearly captures the general
behavior of the data in the intermediate range of $p$. For $p$ close
to zero, one would not expect the system to anticipate the formation
of a gel at $p\approx 0.25$ and for $p$ close to $p_c$ finite-size
effects are clearly evident. Part (b) of this figure attempts to
collapse the data by means of the finite-size scaling ansatz
$\eta(L,p)=L^{s/\nu}\Psi[L(p-p_c)^\nu]$ with $s=0.7$ and
$\nu=0.876$. Internal consistency requires that the scaling function
have the asymptotic form $\Psi(x)\sim x^{-s/\nu}$ for large $x$ and a
line corresponding to this form is also shown on the figure.

\begin{figure}
\centerline{\epsfxsize=6.5cm 
\epsfbox{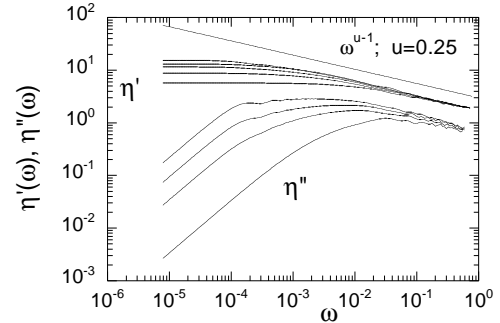}}
\caption{Plot of the complex viscosity $\eta^\ast(\omega)$ as function of
$\omega$ for $L=12$ and $p=0.2$ (lowest curve in each set), $0.22$, $0.23$ and
$0.24$ (top curve in each set). The power law form
$\eta\sim\omega^{u-1}$ is more evident for $\eta'$ than for $\eta''$
and becomes more prominent as $p\to p_c$.}
\label{figG}
\end{figure}

Further support for the conclusion $s\approx 0.7$ comes from the
complex frequency-dependent viscosity $\eta^\ast(\omega)\equiv
\eta'(\omega)+i\eta''(\omega)=G^\ast(\omega)/(i\omega)$ where $G^\ast$
is the complex elastic modulus. The scaling ansatz for these functions
\cite{adam96} is $\lim_{\omega\to 0}\eta'(\omega)\sim (p_c-p)^{-s}$
for $p<p_c$; $\lim_{\omega\to 0}G'(\omega)\sim (p-p_c)^t$ for $p>p_c$
and, for frequencies $\omega >\omega^\ast$, where $\omega^\ast$ is a
characteristic crossover frequency that approaches zero as $p\to p_c$,
$G^\ast(\omega)\sim(i\omega)^u$.  The connection between the critical
behavior of the modulus in the rigid phase and the viscosity in the
fluid phase is then expressed through the scaling relation
$u=t/(s+t)$. Moreover, in the high frequency region, one expects
$\eta'(\omega)$ and $\eta''(\omega)$ to both vary as $\omega^{u-1}$
and the ratio of the real and imaginary parts to obey
$u=2/\pi\tan^{-1}\{\eta'/\eta''\}$.  In our previous work on the rigid
phase \cite{MP99}, we have concluded that $t\approx 2$ in three
dimensions. Therefore, with $s\approx 0.7$ we have $u\approx
0.74$. The frequency-dependent viscosity is plotted in Fig. \ref{figG}
for $L=12$ and $p>0.20$. There is clearly a region of power-law
behavior that extends to lower frequencies as the critical point is
approached. This behavior is seen more clearly in $\eta'$ than in
$\eta''$. Nevertheless, both pieces of the shear viscosity decrease in
a way consistent with $\omega^{-0.27}$ in very satisfactory agreement
with the foregoing analysis. As well, the ratio of $\eta'$ to $\eta''$
in this regime produces a second estimate $u\approx 0.76$.

\begin{figure}
\centerline{\epsfxsize=6.5cm 
\epsfbox{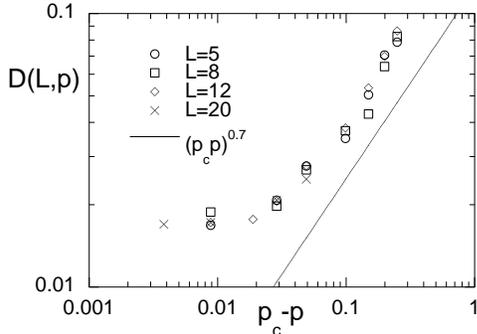}}
\caption{The dimensionless diffusion constant $D(L,p)$ as function of $p$. The
straight line corresponds to a power law $(p_c-p)^{0.7}$.}
\label{figD}
\end{figure}

Finally, in Figure \ref{figD} we show the dimensionless self-diffusion
constant $\sqrt{m/\sigma^2k_BT}D$ obtained from the velocity-velocity
correlator (\ref{D}) for {\it all} particles in the system. In this
case, we show only the raw data. It seems clear from the behavior of
$D$ in the critical region that a finite-size scaling analysis is
unlikely to improve the collapse of the data. For $p <0.2$ the data
are not inconsistent with a power law behavior of the form
$(p_c-p)^{0.7}$ but the evidence for this is weak at best. Moreover,
the fact that the data for $p\approx p_c$ are essentially independent
of $L$ suggests that $D(L\to\infty,p\to p_c)$ is finite. Precisely at
the gel-point, in the thermodynamic limit, the system consists of a
percolating cluster with fractal dimension $D_F\approx 2.5$. The
particles that are not on the infinite cluster are organized into
finite clusters of various sizes. Approximately $18\%$ remain as
monomers that presumably are able to diffuse quite easily through the
percolating cluster since this cluster contains holes on all length
scales. This would account for the absence of critical behavior in
$D$.

We are aware of one previous simulation that attempted to
address the critical behavior of the shear viscosity near the gel
point. Recently del Gado {\it et al.}\cite{dG00} simulated a very
different model, namely particles confined to the sites of a lattice
and randomly crosslinked to form clusters of various sizes. This
system was then evolved by a bond fluctuation method and the diffusion
constants $D_m$ of clusters of mass $m$ measured. They postulated the
relation $D(R)\sim R^{-(1+s/\nu)}$, where $R$ is the radius of
gyration of a cluster, and from this determined $s\approx 1.3$. We are
not aware of any rigorous derivation of this connection between
diffusion and viscosity. However, it may be that their model simply
contains different physics.

In conclusion, we have obtained the
shear viscosity of a viscoelastic liquid below the gel point for a
conceptually simple model that we have previously studied in the rigid
phase. We have presented evidence that the shear viscosity diverges
with an exponent $s\approx 0.7$ consistent with recent theoretical
work\cite{zipp99} and some of the experimental data
\cite{adam}. Future extensions of this work will include an
investigation of viscoelasticity in two dimensions and a study of
dynamics in the gel phase.

One of us (BJ) thanks the Physics
Department at Simon Fraser University, where this work was carried
out, for its hospitality. This work was supported by the NSERC of
Canada.

\end{document}